\documentclass[11pt]{article}
\usepackage{amsmath,amssymb}

\def\a{\alpha}
\def\da{\dot\alpha}
\def\b{\beta}
\def\db{\dot\beta}

\def\d{\delta}

\def\e{\epsilon}           
\def\f{\phi}               
\def\g{\gamma}

\def\j{\psi} \def\bj{\bar\psi}
\def\k{\kappa}                    
\def\l{\lambda}
\def\m{\mu}
\def\n{\nu}
\def\o{\omega}  
\def\q{\theta}    \def\bq{\bar\theta}                
\def\r{\rho}                                     
\def\s{\sigma}                                   

\def\F{\Phi}

\def\O{\Omega}

\def\S{\Sigma}

\def\pa{\partial}              

\begin{document}
\thispagestyle{empty}
\null\vskip-24pt \hfill CERN-TH/2002-056 \vskip-10pt \hfill LAPTH-906/02

\begin{center}
\vskip 0.2truecm {\Large\bf
Conformally coupled supermultiplets in four and five dimensions \vskip
0.5truecm }

{\Large Sergio Ferrara\footnote{Laboratori Nazionali di Frascati, INFN, Italy}
and Emery Sokatchev\footnote{On leave of absence from Laboratoire
d'Annecy-le-Vieux de Physique Th\'{e}orique  LAPTH, B.P. 110, F-74941
Annecy-le-Vieux et l'Universit\'{e} de Savoie}} \vskip 0.4truecm

\vskip 0.3cm {CERN Theoretical Division, CH 1211 Geneva 23, Switzerland}
\end{center}

\vskip 1truecm \Large
\centerline{\bf Abstract} \normalsize We obtain by superfield methods the
exceptional representations of the OSp(2$N/4,\mathbb{R}$) and SU(2,2/1)
superalgebras which extend to supersingletons of SU(2,2/$2N$) and F(4),
respectively. These representations describe superconformally coupled
multiplets and appear in three- and four-dimensional superconformal field
theories which are holographic descriptions of certain anti-de Sitter
supergravities.

\newpage
\setcounter{page}{1}\setcounter{footnote}{0}

\section{Introduction}\label{intro}
In the early days of investigations by physicists \cite{Fronsdal} of the
unitary irreducible representations (UIRs) of SO($d$,2) (the $d$-dimensional
conformal group), it was remarked that there exist special degenerate
representations (``singletons")  which give rise to just a pair of UIRs when
reduced to the subgroup SO($d-1$,2).

These pairs of modules are ``shadow pairs" \cite{FGPG}, i.e., they have the
same Casimir eigenvalues, and correspond to the so-called ``conformally coupled
fields" \cite{Fronsdal} in $d$ dimensions. We recall that SO($d$,2) can be
viewed either as the isometry group of AdS${}_{d+1}$ or as the conformal group
in $d$-dimensional Minkowski space M${}_d$. The two are related by the fact
that a certain compactification of M${}_d$ can be regarded as the boundary of
AdS${}_{d+1}$.

In the present note we extend this analysis to superconformal symmetry. We show
that four-(or five-)dimensional supersingleton multiplets are decomposed in an
analogous way into pairs of three-(or four-)dimensional ``mirror"
supermultiplets corresponding to the superalgebra embedding
OSp($2N/4,\mathbb{R}$) $\subset$ SU(2,2/$2N$) (or SU(2,2/1) $\subset$ F(4)). A
similar result does not hold in the six-dimensional case because the
superalgebra F(4) is not a subalgebra of OSp(8${}^*$/4).

These exceptional representations appeared recently in the physics literature
in different contexts. In the four-dimensional case (with $N=4$) it is the
study of the AdS/CFT correspondence in the presence of ``defects"
\cite{DeWolfe:2001pq,Porrati:2001db}. In the five-dimensional case, in IIB
supergravity on AdS${}_5\times$T${}^{1,1}$ the lowest-dimensional chiral
primary operator ${\rm Tr}(AB)$  describes precisely a conformally coupled
scalar, which means that it has an extension (together with its shadow) to a
supersingleton representation of F(4) \cite{Klebanov:1999tb}.

The paper is organized as follows. In Section 2 we explain how two scalar UIRs
of the $d$-dimensional conformal group SO(d,2) of shadow conformal weights
$\ell_\pm =(d\pm 1)/2$ merge together to form a scalar singleton UIR of the
$d+1$-dimensional conformal group SO(d+1,2). In Section 3 we examine the
possibilities to embed a $d$-dimensional conformal superalgebra in a
$d+1$-dimensional one. We show that OSp($N/4,\mathbb{R}$) $\subset$ SU(2,2/$N$)
and SU(2,2/$1$) $\subset$ F(4), but F(4) $\subset\hskip-9pt/$ OSp($8^*/4$).
Next, in Section 4 we study the branching of $d=4$ scalar supersingletons into
pairs of $d=3$ supermultiplets, corresponding to the embeddings
OSp($1/4,\mathbb{R}$) $\subset$ SU(2,2/$1$) and OSp($2N/4,\mathbb{R}$)
$\subset$ SU(2,2/$2N$). We show that the two $d=3$ ``shadow" scalars obtained
by breaking up the $d=4$ singleton, are found in two ``mirror" $d=3$
supermultiplets of the same conformal dimension 2, but belonging to two
complementary R symmetry irreps (self-dual and anti-self-dual tensors of
SO(2N)). The $d=3$ multiplets obtained can be viewed as ``supercurrents", i.e.,
as squares of the two types of $d=3$ supersingletons. In Section 5 we perform a
similar reduction from $d=5$ to $d=4$, corresponding to the embedding
SU(2,2/$1$) $\subset$ F(4). We show that the $d=5$ supersingleton branches into
a pair of $d=4$ chiral supermultiplets with ``anomalous" dimension 3/2.

\section{Relationship between scalar UIRs of SO($d$,2) and SO($d+1$,2)} \label{sect2}

In this section, following Fronsdal \cite{Fronsdal}, we explain the basic
phenomenon which underlies this work. The question we want to answer is how to
upgrade a scalar\footnote{The same question can be asked about higher spin
representations \cite{Fronsdal}, but scalars are sufficient for our purpose.}
UIR of the conformal group in $d$ dimensions SO($d$,2) to a UIR of the
conformal group in $d+1$ dimensions  SO($d+1$,2). The answer is very simple: We
have to take {\it two UIRs} of SO($d$,2), namely, scalars of conformal
dimension $\ell_1 =(d-1)/2$ and $\ell_2 = (d+1)/2$; together they form a {\it
massless} scalar UIR of SO($d+1$,2) (or a ``singleton" in the AdS sense
\cite{ff2}) of canonical conformal dimension $(d-2)/2$. The explanation
follows.

The algebra of SO($d$,2) contains SO($d-1$,1) (Lorentz) generators
$M_{mn}=-M_{nm}$, translations $P_m$, boosts $K_m$ and dilatation $D$. Here
$m=0,1,\ldots,d-1$ is a vector index of SO($d-1$,1). The commutation relations
relevant to our discussion are
\begin{equation}\label{algebra}
  [P_m,K_n] = iM_{mn} + i\eta_{mn} D \,, \quad [D,P_m]=iP_m\,, \quad
  [D,K_m]=-iK_m\,.
\end{equation}

The group SO($d$,2) is non-compact, so its UIRs are infinite-dimensional
Hilbert spaces. To construct such a space, consider a ground state
$|\ell,0\rangle$ of conformal dimension $\ell$ and spin 0 (i.e., a singlet of
the Lorentz subgroup SO($d-1$,1)) defined by
\begin{equation}\label{grst}
  M_{mn}|\ell,0\rangle = K_m |\ell,0\rangle = 0\,, \quad D|\ell,0\rangle =
  i\ell |\ell,0\rangle\,.
\end{equation}
Here the boosts $K_m$ play the r\^ole of annihilators whereas the translations
$P_m$ are the creators. Applying the latter to the ground state, we obtain the
Hilbert space
\begin{equation}\label{Hilb}
  {\cal H}_{\ell,0}\, : \qquad |\ell,0\rangle\,, \ P_m|\ell,0\rangle\,, \
  P_mP_n|\ell,0\rangle\,, \ \ldots\,.
\end{equation}
In it we find states with spin 0 obtained by acting with the scalar $P^mP_m$:
\begin{equation}\label{scalst}
  |\ell+2k,0\rangle = (P^mP_m)^k|\ell,0\rangle\,.
\end{equation}
Note the absence of states $|\ell+2k+1,0\rangle$ in ${\cal H}_{\ell,0}$.

Now, we wish to upgrade this SO($d$,2) UIR to a UIR of SO($d+1$,2). The algebra
of SO($d+1$,2) contains $d$ additional Lorentz generators $M_{md}$, one boost
(annihilator) $K_d$ and one translation (creator) $P_d$. If $|\ell,0\rangle$ is
to become the ground state of the SO($d+1$,2) UIR with spin 0 (a singlet of
SO($d$,1)) that we want to build up, then it must be annihilated by $M_{md}$
and $K_d$:
\begin{equation}\label{grst2}
  M_{md}|\ell,0\rangle = K_d |\ell,0\rangle = 0\,.
\end{equation}
Further, the new creator $P_d$ applied to this ground state produces the state
\begin{equation}\label{newstate}
  |\ell+1,0\rangle = P_d|\ell,0\rangle\,,
\end{equation}
which is not present in ${\cal H}_{\ell,0}$ (\ref{Hilb}). We must regard this
state as the {\it ground state} of a new SO($d$,2) Hilbert space ${\cal
H}_{\ell+1,0}$. At the same time, it should be viewed as an {\it excitation} of
the SO($d+1$,2) ground state. This explains why we need two UIRs of SO($d$,2)
to form one UIR of SO($d+1$,2).

Next, applying $P_d$ twice we create the state $|\ell+2,0\rangle =
(P_d)^2|\ell,0\rangle$. This time, a similar state is contained in ${\cal
H}_{\ell,0}$ (\ref{Hilb}), namely $P^mP_m|\ell,0\rangle$. Irreducibility then
leads to the identification of these two states:
\begin{equation}\label{KG}
  P^mP_m|\ell,0\rangle = (P_d)^2|\ell,0\rangle \ \Leftrightarrow\
  P^\m P_\mu|\ell,0\rangle= 0\,, \quad \mu=0,1,\ldots,d\,.
\end{equation}
Thus, in $d+1$ dimensions we are dealing with a {\it massless UIR} of the
conformal group SO($d+1$,2) (or a ``singleton" in the AdS terminology).

Finally, the identification condition (\ref{KG}) should be compatible with the
action of all the annihilators $K_\mu$ of SO($d+1$,2). This leads to the
well-known result that the conformal dimension must take the canonical value
for a $d+1$-dimensional massless scalar,
\begin{equation}\label{fixdim}
  P^\m P_\m|\ell,0\rangle= 0 \ \Rightarrow \ \ell=\frac{(d+1)-2}{2} = \frac{d-1}{2}\,.
\end{equation}

In summary, the two SO($d$,2) ground states $|(d-1)/2,0\rangle$ and
$|(d+1)/2,0\rangle$ create two Hilbert spaces ${\cal H}_{(d-1)/2,0}$ and ${\cal
H}_{(d+1)/2,0}$ which merge into the single Hilbert space ${\cal
H}'_{(d-1)/2,0}$ of an SO($d+1$,2) {\it massless} UIR of canonical dimension
$\ell=(d-1)/2$ and Lorentz spin 0.

Note that $\ell_\pm = (d\pm 1)/2$ are the ``shadow dimensions" \cite{FGPG}
corresponding to the same value of the quadratic Casimir of SO($d$,2) (or ``AdS
mass" \cite{Binegar:1982fv,Gubser:1998bc,Witten:1998qj})
$$C_2=M^2 = \ell (\ell-d) = \frac{1}{4}(1-d^2)\;.$$

\section{Embedding of superconformal algebras in diverse dimensions}\label{sect4}

In the preceding section we have seen that two shadow scalars in $d$ dimensions
combine to form a singleton (massless) scalar in $d+1$ dimensions. In the
supersymmetric case we would expect to find a similar relationship between
scalar supersingletons in $d+1$ dimensions (massless scalar multiplets of the
$N$-extended superconformal algebra ${\cal A}^N_{d+1}$) and pairs of scalar
multiplets of the superconformal algebra  ${\cal A}^N_{d}$ in $d$ dimensions.
An obvious necessary condition for this to take place is the existence of the
embedding ${\cal A}^N_{d}\subset {\cal A}^N_{d+1}$.

In this section we discuss the structure of the superconformal algebras
OSp($N/4,\mathbb{R}$) ($d=3$), SU(2,2/$N$) ($d=4$), F(4) ($d=5$) and
OSp($8^*/2N$) ($d=6$) \cite{bgg,ggg}. We show that the embeddings
OSp($N/4,\mathbb{R}$) $\subset$ SU(2,2/$N$) ($d=3\leftarrow d=4$)
\cite{Ferrara:1977sc} and SU(2,2/$1$) $\subset$ F(4) ($d=4\leftarrow d=5$)
exist, but F(4) $\subset$ OSp($8^*/4$)  ($d=5\leftarrow d=6$) is not possible.

To do this, we examine the structure of the anticommutator of the odd
generators case by case, starting from the highest dimension $d=6$ for which a
standard superconformal algebra exists \cite{Nahm:1977tg}.

The $d=6$ superalgebra OSp($8^*/2N$) has the even part ${\rm SO}(6,2)\times
{\rm USp}(2N)$. Correspondingly, the odd generators $\S^a_\a$ carry two
indices. The index $\a=1,\ldots,8$ is in one of the two inequivalent spinor
irreps of SO(6,2) (a non-compact form of SO(8)), e.g., {\bf 8}${}_s$. The index
$a= 1,\ldots,2N$ is in the fundamental irrep of USp(2$N$). These generators
satisfy the pseudoreality condition
\begin{equation}\label{2.1}
  \S^a_\a = C_{\a\b}\O^{ab}(\S^b_{\b})^* \,,
\end{equation}
where $C_{\a\b}$ is the charge conjugation matrix and $\O^{ab}=-\O^{ba}$ is the
USp(2$N$) symplectic matrix. The anticommutator of two odd generators has the
following structure:
\begin{equation}\label{2.2}
  \{\S^a_\a, \S^b_\b\} = \O^{ab} \g^{\m\n}_{\a\b} M_{\m\n} + \mathbb{I}_{\a\b}
  T^{ab}\,.
\end{equation}
Here $M_{\m\n}=-M_{\n\m}$ are the generators of SO(6,2), $\mathbb{I}_{\a\b}$ is
the identity matrix in the appropriate Majorana-Weyl basis and $T^{ab}=T^{ba}$
are the generators of USp(2$N$). We also note the decomposition of a $8\times
8$ matrix $A_{\a\b}$ in terms of antisymmetrized products of gamma matrices:
\begin{equation}\label{2.3}
  A_{\a\b} = \mathbb{I}_{(\a\b)} A^1 + \g^{\m\n}_{[\a\b]} A^{28}_{\m\n} +
  \g^{\m\n\l}_{(\a\b)} A^{35}_{\m\n\l}\,.
\end{equation}
Here we have indicated the symmetry properties of the gamma matrix products, as
well as the dimension of each SO(6,2) irrep.

Now, consider the unique $d=5$ superconformal algebra F(4) \cite{Kac} whose
even part is ${\rm SO}(5,2)\times {\rm USp}(2)$. The odd generators and their
anticommutator have the same form as in the case $d=6$, $\a=1,\ldots,8$ being
an index of the unique spinor representation of SO(5,2) (a non-compact form of
SO(7)) and $a=1,2$ being a USp(2) $\sim$ SU(2) index. What changes, however, is
the decomposition (\ref{2.3}):
\begin{equation}\label{2.4}
  A_{\a\b} = \mathbb{I}_{(\a\b)} A^1 +\g^\m_{[\a\b]} A^7_{\m} + \g^{\m\n}_{[\a\b]} A^{21}_{\m\n} +
  \g^{\m\n\l}_{(\a\b)} A^{35}_{\m\n\l}\,.
\end{equation}
We see that the {\bf 28} of SO(6,2) splits into a {\bf 7} and a {\bf 21} of
SO(5,2), both being antisymmetric matrices. Consequently, trying to break up
the OSp($8^*/4$) anticommutator (\ref{2.2}) so that the corresponding F(4)
relation would emerge, we encounter a problem: It is not possible to keep only
the 21 generators of SO(5,2) without the extra 7 generators of SO(6,2). We
conclude that the embedding F(4) $\subset$ OSp($8^*/4$) does not exist.

The next step is to move from $d=5$ to $d=4$. The $d=4$ superconformal algebra
SU(2,2/$N$) has the even part  ${\rm SO}(4,2)\times {\rm (S)U}(N)$ (in general,
the R symmetry group is U($N$), except for $N=4$ where it is SU(4)). This time
the odd generators $\S^a_\a$ (and their conjugates $\bar\S^\a_a$) carry indices
$\a=1,\ldots,4$ and $a=1,\ldots,N$ in the fundamental irreps of SU(2,2) $\sim$
SO(4,2)
 and of SU($N$), respectively:
\begin{equation}\label{2.5}
  \{\S^a_\a, \bar\S^\b_b\} = \d^b_a (\g^{\m\n})_{\a}^\b M_{\m\n} + \d^\b_\a
  T^{a}_b\,.
\end{equation}
The relevant matrix decomposition now concerns two kinds of matrices:
\begin{equation}\label{2.6}
  A_{\a\b} = \g^\m_{[\a\b]} A^6_{\m} + \g^{\m\n\l}_{(\a\b)} A^{10}_{\m\n\l}\,, \qquad
  A_\a^\b = \d_\a^\b A^1 + (\g^{\m\n})_{\a}^\b A^{15}_{\m\n}\;,
\end{equation}
where $(\g^{\m\n})_{\a}^\b$ is traceless. Let us now try to reduce F(4) to
SU(2,2/1) (we set $N=1$ in order to match the numbers of odd generators). To
this end we have to break up the $d=5$ generators $\S^a_\a$ into a pair of
$d=4$ generators $\S_\a$ and $\bar\S^\b$ and then try to fit the pieces of the
generators of the even part. We note that the {\bf 21} of SO(5,2) decomposes
into {\bf 15} and {\bf 6} of SO(4,2). According to (\ref{2.6}), the symmetry
property of the {\bf 6} does not match that of the $d=4$ anticommutator
$\{\S_\a,\S_\b\}$, so we have to set the latter to zero. Further, there is room
for the {\bf 15} in the $d=4$ anticommutator $\{\S_\a,\bar\S^\b\}$. Thus, we
arrive at the SU(2,2/1) algebra:
\begin{equation}\label{2.7}
  \{\S_\a,\S_\b\} = 0\,, \qquad \{\S_\a,\bar\S^\b\} = (\g^{\m\n})_{\a}^\b
  M_{\m\n}+ \d_{\a}^\b T\,.
\end{equation}
We conclude that the embedding SU(2,2/1) $\subset$ F(4) is possible.

Finally, consider the reduction from $d=4$ to $d=3$. The superalgebra
OSp($N/4,\mathbb{R}$) has the even part ${\rm SO}(3,2)\times {\rm SO}(N)$. The
odd generators $\S^a_\a$ carry indices $\a=1,\ldots,4$ in the spinor irrep of
Sp($4,\mathbb{R}$) $\sim$ SO(3,2) and $a=1,\ldots,N$ in the vector irrep of
SO($N$). In addition, they satisfy the reality condition $\S^a_\a =
\O_{\a\b}\bar\S^{a\b}$. Their anticommutator is
\begin{equation}\label{2.8}
  \{\S^a_\a, \S^b_\b\} = \d^{ab} (\g^{\m\n})_{\a\b}M_{\m\n}+ \O_{\a\b}
  T^{ab}\;,
\end{equation}
where $T^{ab}=-T^{ba}$ are the SO($N$) generators. Note also the matrix
decomposition
\begin{equation}\label{2.9}
  A_{\a\b} = \O_{\a\b} A^1 + \g^\m_{[\a\b]} A^5_{\m} + \g^{\m\n}_{(\a\b)} A^{10}_{\m\n}\,,
\end{equation}
where $\g^\m_{[\a\b]}$ is traceless, $\O^{\a\b}\g^\m_{[\a\b]}=0$. Now we can
see how the reduction from SU(2,2/$N$) works: The $d=3$ odd generators are the
real parts of the $d=4$ ones; U($N$) becomes SO($N$); SU(2,2) becomes
Sp($4,\mathbb{R}$). At the last step the {\bf 15} of SU(2,2) breaks up into the
{\bf 10} and the {\bf 5} of Sp($4,\mathbb{R}$), the former having the right
symmetry property to fit in (\ref{2.8}), the latter has to drop out.

\section{The reduction of SU(2,2/$N$) scalar supersingletons to two
OSp($N/4,\mathbb{R}$) mirror multiplets}\label{sect3}

In this section we explain the mechanism of the decomposition of the $d=4$
scalar supersingletons (massless scalar multiplets of SU(2,2/$N$)) into a pair
of scalar multiplets of the $d=3$ superconformal algebra ${\rm
OSp}(N/4,\mathbb{R})\subset {\rm SU}(2,2/N)$.

We start with the simplest case $N=1$ which illustrates the origin of the
shadow scalars in $d=3$. The $d=4$ supersingleton multiplet consists of
massless scalars and spinors. It is described by an ultrashort superfield
(chiral in the case $N=1$). Its $\q$ expansion contains a term with first-order
derivatives of the scalars. Unlike standard trivial dimensional reduction to
$d=3$, our procedure consists in keeping the derivative $\pa_3$ of the scalars.
It plays the r\^ole of the additional creation operator $P_3$ described in
Section \ref{sect2} and is thus responsible for the origin of the shadow
scalars.

Next we discuss in detail the case with $N=2$ extended supersymmetry, as well
as the generalization to an arbitrary even value of $N$. This time the $d=4$
supersingletons are not chiral, but Grassmann analytic ultrashort superfields
in harmonic superspace \cite{Galperin:1984av,Howe:md,Galperin:uw}. Here we see
a new feature: The scalars in $d=4$ form an irrep $R_4$ of the $d=4$ R symmetry
group SU($N$) which splits into two inequivalent halves under the $d=3$ group
SO($N$), $R_4=R^+_3 + R^-_3$. Accordingly, the $d=4$ supersingleton decomposes
into two ``mirror" $d=3$ supermultiplets. One of them has scalars of dimension
1 in, e.g., $R^+_3$ as its ``ground state" and also contains scalars of
dimension 2 in $R^-_3$. The latter play the r\^ole of the ``shadows" of the
ground state scalars of the mirror multiplet.

Another peculiarity of the reduction SU(2,2/$N$) $\rightarrow$
OSp($N/4,\mathbb{R}$) is that the $d=3$ supermultiplets can be viewed as
``supercurrents", i.e., as the squares of the $d=3$ scalar supersingletons.
This implies that such representations correspond to ``massless" fields in
AdS${}_4$ \cite{ff2,Witten:1998qj,Ferrara:1998jm}. For $N\geq 2$ there are two
inequivalent species of $d=3$ scalar supersingletons. Each of them gives rise
to one of the mirror supercurrent multiplets obtained by decomposing the $d=4$
scalar supersingleton.

\subsection{$N=1$}

The $d=4$ $N=1$ scalar supersingleton (the massless Wess-Zumino multiplet) is
described by a chiral superfield $W(x,\theta,\bar\theta)$ subject to the
massless field equation:
\begin{equation}\label{1.1}
  \bar D_{\dot\alpha} W = D^\alpha D_\alpha W = 0\,.
\end{equation}
The component expansion of this superfield has the form
\begin{equation}\label{1.2}
  W = \phi(x) + \theta^\alpha \psi_\alpha(x) + i\theta^\alpha
  \sigma^\mu_{\alpha\dot\alpha}\bar\theta^{\dot\alpha} \partial_\mu \phi(x)\,,
\end{equation}
where the complex scalar $\phi(x)$ and the left-handed spinor $\psi_\alpha(x)$
fields are massless:
\begin{equation}\label{1.3}
  \square \phi = \tilde\sigma_\mu^{\dot\alpha\alpha} \partial^\mu
  \psi_\alpha = 0\,.
\end{equation}
Note that the space-time coordinates $x^\mu$ in (\ref{1.2}) are real. The
derivative term originates from the nilpotent shift defining the complex chiral
basis $x^\mu_L = x^\mu + i\theta^\alpha
\sigma^\mu_{\alpha\dot\alpha}\bar\theta^{\dot\alpha} $, in which $W$ depends on
$\theta$ only. This property is called ``chirality", which is a particular case
of ``Grassmann analyticity"(or ``1/2 BPS shortness"). In fact, the superfield
(\ref{1.2}) is even ``ultrashort", since the term $\q^\a\q_\a$ is missing. This
is a characteristic feature of the supersingleton superfields.

The reduction to three dimensions is achieved by identifying the two conjugate
Grassmann variables
\begin{equation}\label{1.4}
   \theta_\alpha =  \bar\theta_{\dot\alpha}
\end{equation}
and by taking, e.g., the real part of the superfield $W$, $J= \mbox{Re}\; W$.
We stress the important difference between the procedure we follow here and the
standard trivial dimensional reduction. In the latter case we would also set
$\partial_3 \phi = 0$. Then, since the $d=4$ sigma matrices
$\sigma^\mu_{\alpha\dot\alpha}$ split into the $d=3$ ones
$\sigma^m_{(\alpha\beta)}$ (symmetric) and $\sigma^3_{\alpha\beta} =
i\epsilon_{\alpha\beta}$, the derivative term in (\ref{1.2}) would drop out and
we would get the short real $d=3$ superfield $J=A+i\theta^\alpha
\lambda_\alpha$, where $A = \mbox{Re}\; \phi$, $\lambda = \mbox{Im}\; \psi$.
Such a superfield satisfies the superspace constraint $D^\alpha D_\alpha J =
0$, which is easily shown to be superconformal only if the conformal dimension
of the superfield $J$ equals 1/2 and not 1 (the dimension of $W$). This is an
indication that we have to proceed differently.

The correct dimensional reduction procedure which preserves conformal
invariance, consists in keeping $\partial_3 \phi$. According to the discussion
in Section \ref{sect2}, this corresponds to considering $\pa_3$ as the
component of the raising operator $P_3$ which is responsible for creating a new
scalar state. In terms of fields this means defining a new $d=3$ field $B = -
\mbox{Im}\;
\partial_3 \phi$. Thus, we obtain a generic long real\footnote{This is the
reason why we had to expand the chiral superfield $W$ (\ref{1.2}) in the real
basis in superspace, and not in the complex chiral basis.} $d=3$ superfield:
\begin{equation}\label{1.5}
  J=A(x)+i\theta^\alpha
\lambda_\alpha(x) + i\theta^\alpha \theta_\alpha B(x)\;,
\end{equation}
$$A = \mbox{Re}\; \phi\;, \qquad \lambda_\alpha =
\mbox{Im}\; \psi_\alpha\;, \qquad B = - \mbox{Im}\; \partial_3 \phi\;.$$ In the
process the first scalar $A$ keeps the original conformal dimension 1 of
$\phi$, but the new scalar $B$ acquires the ``shadow" dimension $3-1=2$.

Instead of the real part of $W$, we could have taken its imaginary part, thus
obtaining an alternative (``mirror") $d=3$ supermultiplet:
\begin{equation}\label{1.6}
  J'=A'(x)+i\theta^\alpha
\lambda'_\alpha(x) + i\theta^\alpha \theta_\alpha B'(x)\;,
\end{equation}
$$A' = \mbox{Im}\; \phi\;, \qquad \lambda'_\alpha =
-\mbox{Re}\; \psi_\alpha\;, \qquad B' =  \mbox{Re}\; \partial_3 \phi\;.$$ The
only difference between $J$ and $J'$ has to do with the fact that real part of
$\phi$ is a scalar while its imaginary part is a pseudoscalar. Later on we
shall see that the difference between the mirror $d=3$ multiplets becomes more
significant for $N>1$.

Finally, $J$ can be viewed as a ``supercurrent" obtained from $d=3$
supersingletons $\Phi(x,\theta)$ defined by the massless superfield equation
\begin{equation}\label{1.7}
  D^\alpha D_\alpha \Phi = 0\ \Rightarrow \ \Phi=\omega(x) + \theta^\alpha
  \chi_\alpha \;, \qquad \square \omega = \partial^{\alpha\beta}\chi_\beta =
  0\,.
\end{equation}
Note that this time the dimensions of the $d=3$ fields are canonical,
$\mbox{dim}\;\omega = 1/2$, $\mbox{dim}\;\chi = 1$, so that equation
(\ref{1.7}) can be superconformal. Then we can write the ``supercurrent" $J =
(\Phi)^2$ which has the same content as (\ref{1.5}). The term ``supercurrent"
will become more clear when we move to $N>2$, where we will find conserved
currents among the components of $J$.

\subsection{$N=2$}\label{HM}

The $d=4$ $N=2$ scalar supersingleton is the hypermultiplet. Its adequate
description as a Grassmann analytic superfield is given in harmonic superspace
\cite{Galperin:1984av}. To this end one introduces harmonic variables $u^I_i$
(and their complex conjugates $u^i_I$) parametrizing the coset SU(2)/U(1). They
form a matrix of the R symmetry group SU(2) whose index $i$ transforms under
the fundamental representation of SU(2), whereas the index $I$ is a collection
of U(1) charges. With the help of the harmonics we can covariantly project any
SU(2) representation with respect to the subgroup U(1). For instance, the
Grassmann variables $\bar\theta^i_{\dot\alpha}$ and $\theta_{i\alpha}$ in the
fundamental irrep of SU(2) can be projected onto the highest-weight states
(HWS) $\bar\theta^1_{\dot\alpha} = \bar\theta^i_{\dot\alpha} u^1_i$,
$\theta_{2\alpha} = \theta_{i\alpha} u^i_2$, etc. Note the existence of a
particular conjugation (real structure) which combines ordinary complex
conjugation with the antipodal map on the sphere $S^2\sim\mbox{SU(2)/U(1)}$:
\begin{equation}\label{1.7'}
  \widetilde{\theta}_{2\alpha} =  \bar\theta^1_{\dot\alpha}\,.
\end{equation}

The on-shell hypermultiplet can now be described by a Grassmann analytic
superfield:
\begin{equation}\label{1.8}
  W^1(x_A,\bq^1,\q_2,u) = \f^i(x_A) u^1_i + \q^\a_2
  \j_\a(x_A) + \bq^{1\da}\bar\k_{\da}(x_A) + i \q_2\s^\m \bq^1 \pa_\m \f^i(x_A)
  u^2_i\,,
\end{equation}
where the component fields are massless,
\begin{equation}\label{1.8'}
  \square \f^i = \tilde\pa^{\da\a}\j_\a
= \pa_{\a\da}\bar\k^{\da}=0\,.
\end{equation}
The space-time variables $x^\m_A$ in (\ref{1.8}) are obtained from the real
ones $x^\m$ by a nilpotent shift, $x^\m_A = x^\m +i\q_2\s^\m\bq^2 -
i\q_1\s^\m\bq^1$. In this basis the Grassmann analytic superfield manifestly
depends only on half of the odd variables. In this respect the Grassmann
analytic superfields resemble the chiral ones, although the half of $\q$'s they
depend on is chosen with regard to the R symmetry instead of the Lorentz group.
We can also say that this is another type of 1/2 BPS short superfield.

In addition to Grassmann analyticity, $W^1$ is subject to the constraint of
SU(2) irreducibility (harmonic analyticity), which makes the harmonic
dependence in (\ref{1.8}) linear and puts the physical fields on shell. As in
the case $N=1$, the supersingleton (\ref{1.8}) is an ``ultrashort" superfield,
in the sense that a number of components in its $\q$ expansion are missing.

One final remark concerns the derivative term $i \q_2\s^\m \bq^1 \pa_\m
\f^i(x_A)u^2_i$. It is not the analog of the coordinate shift term in  the
$N=1$ chiral superfield (\ref{1.2}), but is due to the ultrashortness of the
supersingleton. This term will play an important r\^ole in the dimensional
reduction.

Now we proceed to the reduction to three dimensions. There the $N=2$ R symmetry
group SU(2) becomes SO(2) and the Grassmann variables belong to the vector
representation, $\q^{\pm\pm\a}$.\footnote{We use spinor units of charge, so the
charge of a vector is $\pm 2$.} Thus, to reduce the $d=4$ superfield
(\ref{1.8})  we have to identify the two conjugate (in the sense of eq.
(\ref{1.7'})) $\q$'s,
\begin{equation}\label{1.8''}
  \q_2^\a = \bq^{1\da} \equiv \q^{++\a}\,.
\end{equation}
Further, we define the scalar field $A^{++} \equiv \f^i u^1_i$ and the spinor
field $\l^\a \equiv 1/2(\j^\a +\bar\k^{\da})$. Finally, much like in the case
$N=1$, from the derivative term in (\ref{1.8}) only the component $B^{--}
\equiv -\pa_3 \f^i u^2_i$ survives, giving rise to a new scalar field of shadow
dimension 2. The result is
\begin{equation}\label{1.9}
  J^{++} = A^{++} + \q^{++\a} \l_\a + \q^{++\a}\q^{++}_\a B^{--}\,.
\end{equation}
Unlike the case $N=1$, this time we obtain a 1/2 BPS short (or Grassmann
analytic) superfield depending only on half of the odd variables.\footnote{The
space-time variables in (\ref{1.9}) include the nilpotent shift $x^m_A = x^m +
i\q^{++}\s^m\q^{--}$ which originates from the corresponding shift in $d=4$.}
However, this short superfield is off shell, unlike its $d=4$ counterpart
(\ref{1.8}).

A characteristic feature of the $d=3$ supermultiplet (\ref{1.9}), common for
all even values of $N$, is that its ``ground state" $A^{++}$ corresponds to
half of the $d=4$ scalars $\f^i$. It is obtained by splitting the R symmetry
$n$-fold antisymmetric irrep of SU(2$n$) into two inequivalent irreps
(self-dual and anti-self-dual) of SO(2$n$). The other half reappears at the
level of $(\q)^2$, but this time with shadow dimension 2.

Choosing the other half of the scalars as the ``ground state" of the
supermultiplet, we find an alternative (``mirror") reduction of the $d=4$
supersingleton. It is obtained from the complex conjugate superfield
$\overline{W}_2(\q_1, \bq^2) = \bar\f_i u^i_2 + \ldots$ by the identifications
$\q_1^\a = \bq^{2\da} \equiv \q^{--\a}$, $A^{--} = \bar\f_i u^i_2$, $\r^\a
\equiv 1/2(\j^\a -\bar\k^{\da})$ and $B^{++} \equiv -\pa_3 \bar\f_i u_2^i$:
\begin{equation}\label{1.10}
  J^{--} = A^{--} + \q^{--\a} \r_\a + \q^{--\a}\q^{--}_\a B^{++}\,.
\end{equation}
Notice the important fact that the ``shadow" scalar $B^{++}$ in (\ref{1.10})
carries the same R symmetry quantum numbers as the ``ground state" scalar
$A^{++}$ in (\ref{1.9}). In other words, the pair of $d=3$ scalars which gives
rise to a singleton scalar representation of SO(4,2) (recall Section
\ref{sect2}), is composed of states belonging to the two ``mirror" $d=3$
supermultiplets (\ref{1.9}) and (\ref{1.10}). We conclude that this pair of
$d=3$ supermultiplets is equivalent to the $d=4$ supersingleton multiplet
(\ref{1.8}). As we show below, the same phenomenon takes place for all even
values of $N$.

Finally, we remark that the $d=3$ $N=2$ supermultiplet we have obtained is a
``supercurrent", i.e., it can be viewed as the square of a $d=3$
supersingleton. The latter is described by an ultrashort analytic superfield,
\begin{equation}\label{1.11}
  \F^+ = \o^+ + \q^{++\a}\chi^-_\a\,, \quad \square\o^+ = \pa^{\a\b}\chi^-_\b
  = 0\ \Rightarrow\ J^{++} = (\F^+)^2\,,
\end{equation}
or, alternatively, by $\F^- = \o^- + \q^{--\a}\chi^+_\a$, $J^{--} = (\F^-)^2$.

\subsection{Arbitrary even $N$}

The above discussion is easily generalized to the case of an arbitrary even
value $N=2n$. Two examples of $d=4$ supersingletons of physical interest are
the on-shell multiplets of $N=4$ super Yang-Mills and of $N=8$ supergravity
\cite{Gunaydin:vz}.

\subsubsection{Scalar supersingletons in $d=4$}

The $d=4$ $N=2n$ scalar supersingleton is described by an ultrashort analytic
superfield. It depends on half of the Grassmann variables:
\begin{equation}\label{1}
  \bq^{1,2,\cdots,n\; \da} = \bq^{i\da}u^{1,2,\cdots,n}_i\,, \qquad \q^\a_{n+1,\cdots,
  2n}= \q^\a_i u^i_{n+1,\cdots, 2n} \,,
\end{equation}
obtained with the help of the harmonic variables $u^I_i$ (and their conjugates
$u^i_I$) on the coset SU(2$n$)/$[\mbox{U}(1)]^{2n-1}$
\cite{Andrianopoli:1999vr}. They form an SU($2n$) matrix and are used to
project the fundamental irrep of SU($2n$) (index $i$) onto $[{\rm
U}(1)]^{2n-1}$ (index $I=1,2,\ldots,2n$).

Below we list some of the terms in the $\q$ expansion of the supersingleton:
\begin{eqnarray}
 W^{12\cdots n}(\bq^{1,2,\cdots,n}, \q_{n+1,\cdots, 2n}) &=&
 \f^{[i_1\cdots i_n]}u^1_{i_1}\cdots u^n_{i_n}  \nonumber\\
  &+& \bq^{n\;\da}\bj^{[i_1\cdots i_{n-1}]}_{\da}u^1_{i_1}\cdots u^{n-1}_{i_{n-1}}
   + \mbox{perm.} \nonumber\\
  &+& \q_{2n}^\a \k_{[i_1\cdots i_{n-1}]\;\a} u^{i_1}_{2n-1} \cdots
   u_{n+1}^{i_{n-1}}  + \mbox{perm.} \nonumber\\
  &+& \ldots\nonumber\\
  &+& \bq^1_{\da_1}\cdots \bq^n_{\da_n} F^{(\da_1\cdots \da_n)} +
  \q^{\a_1}_{2n}\cdots \q^{\a_n}_{n+1} G_{(\a_1\cdots\a_n)}  \nonumber\\
  &+& i\q_{2n}\s^\m\bq^n
  \pa_\m \f^{[i_1\cdots i_n]}u^1_{i_1}\cdots u^{n-1}_{i_{n-1}}
  u^{2n}_{i_{n}} + \mbox{perm.} \nonumber\\
  &+& \mbox{other derivative terms}
  \label{1.20}
\end{eqnarray}
These include: the scalars in the $n$-fold antisymmetric irrep $[0 \cdots 0
1_n0 \cdots 0_{2n-1} ]$ of SU(2$n$); the spin 1/2 fields in the $n-1$-fold
antisymmetric irrep $[0 \cdots 0 1_{n-1}0 \cdots 0_{2n-1} ]$; the SU(2$n$)
singlets $F$ and $G$ with the top Lorentz spins $(0,n)$ and $(n,0)$; the
derivative terms relevant to the dimensional reduction. All of these fields are
massless,
\begin{equation}\label{1.21}
  \square\f = \pa_{\a\da}\bj^{\da}= \tilde\pa^{\da\a}\k_\a =
  \ldots = \pa_{\a\da_1} F^{(\da_1\cdots
  \da_n)}= \tilde\pa^{\da\a_1}G_{(\a_1\cdots\a_n)} = 0\,.
\end{equation}
The harmonic projections of the fields displayed in (\ref{1.20}) correspond to
the HWS of each SU(2$n$) irrep, the rest are obtained by permutation of the
indices $1,2,\ldots,2n$. The conjugation rules compatible with Grassmann
analyticity are as follows:
\begin{equation}\label{1.21'}
 \widetilde{\q}_{n+1}=\bq^1,\ \ldots,\ \widetilde{\q}_{2n}= \bq^{n}\, .
\end{equation}
Using these rules, we can impose a reality condition on the multiplet
(\ref{1.20}) if $N=2n=4k$.

To be more explicit, let us give the particular example of the $N=4$ SYM
multiplet in detail:
\begin{eqnarray}
  W^{12}_{N=4\ {\rm SYM}}(\bq^{1,2},\q_{3,4})&=&\f^{[ij]} u^1_i u^2_j \label{1.12}\\
  &+& (\bq^{2\da}\bj^i_{\da} u^1_i
   - 1 \leftrightarrow 2) + (\q^\a_4 \j_{\a i}u^i_3 - 3 \leftrightarrow 4)   \nonumber\\
  &+& \bq^{1\da} \bq^{2\db} F^-_{(\da\db)} + \q_4^\a \q_3^\b F^+_{(\a\b)}   \nonumber\\
  &+& \left((i\q_4\s^\m\bq^2 \pa_\m \f^{[ij]} u^1_i u^4_j + 3 \leftrightarrow 4)
  - 1 \leftrightarrow 2\right) + \mbox{other derivative terms}\,. \nonumber
\end{eqnarray}
It consists of six scalars $\f^{ij} = 1/2\;\e^{ijkl}\bar\f_{kl}$ in the real
irrep $[010]$ of SU(4), four spinors $\bj^i_{\da}$ in the fundamental irrep
$[100]$ (and their conjugates $\j_{\a i}$ in $[001]$) and the self-dual
$F^+_{(\a\b)}$ and anti-self-dual $F^-_{(\da\db)}$ parts of the YM field
strength (SU(4) singlets).

In eqs. (\ref{1.20}) and(\ref{1.12}) we have displayed only one of the numerous
derivative terms. The remainder contains either derivatives of the spinors or
higher-order derivatives of the scalars. They are irrelevant for the
dimensional reduction because of the field equations (\ref{1.21}). Indeed, a
term like $\pa_\mu\j_\a$ will only generate derivative terms (descendants) in
$d=3$, since $\pa_3 \j_\a = \tilde\s_3^{\a\da}\s_{\da\b}^m \pa_m \j^\b$.
Similarly, since $\pa_3^2\f = \pa^2_m\f$, derivative terms like $\pa_\m \pa_\n
\f$ either give rise to descendants of the $d=3$ field $A\equiv\f$ or of the
new $d=3$ field $B\equiv\pa_3\f$ (see below).

\subsubsection{Reduction to $d=3$}

By analogy with the case $N=2$, after the reduction to $d=3$ we expect to find
an ${\rm SO}(2$n$)$ covariant analytic superfield. In order to define it we
have to introduce harmonic variables on the coset  SO(2$n$)/$[\mbox{U}(1)]^n$.
\footnote{We did not need harmonics in the case $N=2$ since the group SO(2) is
too small to admit a harmonic coset.} A characteristic feature of the
orthogonal groups SO(2$n$) (more precisely, of their covering groups
Spin(2$n$)) is the existence of two inequivalent spinor representations (left-
and right-handed), denoted by undotted and dotted indices $a,\dot a =
1,\ldots,2^{n-1}$. Correspondingly, we introduce two sets of spinor harmonics,
\begin{equation}\label{1.12'}
  u^{\pm_1\cdots\pm_{n-2}[\pm]}_a\,, \qquad
  w^{\pm_1\cdots\pm_{n-2}\{\pm\}}_{\dot a}\,.
\end{equation}
Their U(1) charges denoted by $\pm_p$, $p=1,\ldots,n-2$ correspond to the first
$n-2$ U(1) factors from the coset SO(2$n$)/$[\mbox{U}(1)]^n$. The last two
charges, denoted by $[\pm]$ and $\{\pm\}$, are used to distinguish the two
spinor representations whose HWS correspond to the projections
\begin{equation}\label{1.22}
 [0 \cdots 0 1 0_n]\,:\  u^{+_1\cdots+_{n-2}[+]}_a\,;
 \qquad [0 \cdots 0 1_n] \,:\  w^{+_1\cdots+_{n-2}\{+\}}_{\dot
 a}\,.
\end{equation}

Further, in $d=4$ the Grassmann variables $\q_i$ and $\bq^i$ belong to the
fundamental irrep of SU(2$n$), but after reducing to $d=3$ and taking the real
part, we find them in the vector irrep of SO(2$n$). The corresponding vector
harmonics are composite variables made out of the left- and right-handed spinor
harmonics,
\begin{equation}\label{1.14}
  v^R_r  =
  u^{\pm_1\cdots\pm_{n-2}[\pm]}_a\; \gamma_r^{a\dot a}\;
  w^{\pm_1\cdots\pm_{n-2}\{\pm\}}_{\dot a}\,,
\end{equation}
where $\gamma_r$, $r=1,\ldots,2n$ are the gamma matrices of SO(2$n$). The
projection (charge) $R$ takes one of the following $2n$ values: $\pm\pm_1$ or
$\pm\pm_2$, \ldots, or $\pm\pm_{n-2}$, or $[\pm]\{\pm\}$. The HWS of the vector
irrep $[10\cdots0_n]$ corresponds to the projection $v^{++_1}_r$ (if $n>2$) or
to $v^{[+]\{+\}}$ (if $n=2$).

Using the composite vector harmonics, we can choose a subset of the odd
variables suitable for defining a $d=3$ analytic superspace. One such subset is
obtained by identifying the left- and right-handed $d=4$ analytic $\q$'s as
follows:
\begin{equation}\label{1.24}
  \q_{n+1}=\bq^1\equiv\q^{++_1},\ \ldots,\
  \q_{2n-2}=\bq^{n-2}\equiv\q^{++_{n-2}},\
  \q_{2n-1}= \bq^{n-1}\equiv \q^{[+]\{+\}},\  \q_{2n}= \bq^{n}\equiv
  \q^{[+]\{-\}}\, .
\end{equation}
These new odd variables are real, as follows from the conjugation rules
(\ref{1.21'}). Another analytic subset  is obtained by swapping
$\q^{[+]\{\pm\}}\leftrightarrow\q^{[\pm]\{+\}}$.

The next step in the reduction is to split the scalars in eq. (\ref{1.20}) into
irreps of SO(2$n$). In fact, all other fields in  (\ref{1.20}) belong to
SU(2$n$) irreps of the type $[0 \cdots 0 1_p 0 \cdots 0_{2n-1}]$, $1\leq p \leq
n-1$, which remain irreducible when restricted to SO(2$n$) (more precisely, to
Spin($2n$)). The only exception are the scalars in the $[0 \cdots 0  1_n 0
\cdots 0_{2n-1}]$ of SU(2$n$) which split into the left- and right-handed
irreps $[0 \cdots 0  20_n]$ and $[0 \cdots 0 2_n]$ of SO(2$n$) (corresponding
to self-dual and anti-self-dual $n$-fold antisymmetric tensors). We can take,
e.g., the self-dual half $[0 \cdots 0  20_n]$ (i.e., the HWS charges
$++_1\cdots++_{n-2}[++]$) and define the $d=3$ ``ground state" scalar fields
\begin{equation}\label{1.16}
 \f^{[i_1\cdots i_n]} u^1_{i_1}\cdots u^n_{i_n} \ \rightarrow\
 A^{++_1\cdots++_{n-2}[++]} =
 A^{[r_1 \cdots r_n]} v^{++_1}_{r_1}\cdots v^{++_{n-2}}_{r_{n-2}}
 v^{[+]\{+\}}_{r_{n-1}}v^{[+]\{-\}}_{r_{n}}\,.
\end{equation}
Here $A^{[r_1 \cdots r_n]}$ is an $n$-fold antisymmetric self-dual tensor of
SO(2$n$). If $n=2k$ it can be made real by applying the appropriate harmonic
conjugation.

In a similar fashion, the remaining terms in the $d=4$ expansion (\ref{1.20})
give rise to $d=3$ fields. For instance, the spin 1/2 terms become
\begin{equation}\label{1.16'}
  \bq^{n\da}\bj^{12\cdots n-1}_{\da} + \q^\a_{2n} \k_{2n-1\cdots n+1\; \a} \
  \rightarrow  \ \q^{[+]\{-\}\a} \l^{++_1\cdots++_{n-2}[+]\{+\}}_\a \,,
\end{equation}
where $\l^{++_1\cdots++_{n-2}[+]\{+\}}_\a = \l_{\a\; [r_1\cdots r_{n-1}]}
v^{++_1}_{r_1}\cdots v^{++_{n-2}}_{r_{n-2}} v^{[+]\{+\}}_{r_{n-1}}$  is a
spinor filed corresponding to the HWS of the $n-1$-fold antisymmetric irrep $[0
\cdots 0 11_n]$ of SO(2$n$).

This process goes on until we reach the top spin fields:
\begin{equation}\label{topspin}
  \bq^1_{\da_1}\cdots \bq^n_{\da_n} F^{(\da_1\cdots \da_n)} +
  \q^{\a_1}_{2n}\cdots \q^{\a_n}_{n+1} G_{(\a_1\cdots\a_n)}  \
  \rightarrow  \ \q^{++_1}_{\a_1}\cdots \q^{++_{n-2}}_{\a_{n-2}}
  \q^{[+]\{+\}}_{\a_{n-1}} \q^{[+]\{-\}}_{\a_{n}} j^{(\a_1\cdots\a_n)}\,.
\end{equation}
Note that the spin $n$ filed $j^{(\a_1\cdots\a_n)}$ (as well as all other
fields with spin greater than 1/2) is a conserved spin-tensor:
\begin{equation}\label{spten}
  \s_{\a_1\a_2}^m \pa_m j^{(\a_1\cdots\a_n)} = 0\,.
\end{equation}
The reason for this will become clear in subsection \ref{singleton}.

The derivative term shown in (\ref{1.20}) deserves particular attention,
because it gives rise to the new scalars of shadow dimension 2:
\begin{equation}\label{1.16''}
  i\q_{2n}\s^3 \bq^{n}\; \pa_3 \f^{1\cdots n-1; 2n}\ \rightarrow \
  \q^{[+]\{-\}\a}\q^{[+]\{-\}}_\a B^{++_1\cdots++_{n-2}\{++\}} \,,
\end{equation}
where $B^{++_1\cdots++_{n-2}\{++\}} = B_{[r_1 \cdots r_n]} v^{++_1}_{r_1}\cdots
v^{++_{n-2}}_{r_{n-2}} v^{[+]\{+\}}_{r_{n-1}} v^{[-]\{+\}}_{r_{n}}$ is a scalar
filed corresponding to the HWS of the $n$-fold antisymmetric anti-self-dual
irrep $[0 \cdots 0 2_n]$.

As in the case $N=2$, the ``ground state" scalars $A$ and the ``shadow" scalars
$B$ belong to the two complementary SO(2$n$) irreps $[0 \cdots 0 20_n]$ and $[0
\cdots 0 2_n]$ which constitute the SU(2$n$) irrep $[0 \cdots 0  1_n 0 \cdots
0_{2n-1}]$. This means that the shadows of the $A$ fields (\ref{1.16}) are not
the $B$ fields (\ref{1.16''}) from the analytic superfield $J^{[0 \cdots 0 2
0_n]}(\q^{++_1},\ldots,\q^{++_{n-2}},\q^{[+]\{\pm\}})$ that we have
constructed. They can be found in the ``mirror" superfield $J^{[0 \cdots 0
2_{n}]}(\q^{++_1},\ldots,\q^{++_{n-2}},\q^{[\pm]\{+\}})$ obtained by swapping
the irreps of the ``ground state" and of the shadow scalars (in practice, this
means swapping the charges $[\pm]$ and $\{\pm\}$). We conclude that the $d=4$
supersingleton decomposes into two ``mirror" $d=3$ multiplets, each of them
containing the shadow scalars for the other.

Concluding this subsection, we give as an example the complete expansion of the
$d=3$ $N=4$ analytic superfield obtained from the $d=4$ $N=4$ SYM
supersingleton:
\begin{eqnarray}
  J^{[++]}_{N=4\ {\rm SYM}}&=& A^{[++]}  \nonumber\\
  &+&   \q^{[+]\{-\}\a}\l^{[+]\{+\}}_\a -
   \q^{[+]\{+\}\a}\l^{[+]\{-\}}_\a  \nonumber\\
  &+& \q^{[+]\{-\}\a} \q^{[+]\{-\}}_\a B^{\{++\}}
  - \q^{[+]\{-\}\a} \q^{[+]\{+\}}_\a B^{\{-+\}} +
  \q^{[+]\{+\}\a} \q^{[+]\{+\}}_\a B^{\{--\}}\nonumber\\
  &+& \q^{[+]\{+\}\a}\s^m_{\a\b}\q^{[+]\{-\}\b} j_m\nonumber\\
  &+& \mbox{derivative terms} \label{1.18}
\end{eqnarray}
where the vector $j_m$ is conserved, $\pa^m j_m =0$.

\subsection{Supercurrent multiplets}\label{singleton}

In the cases $N=1,2$ we have seen that the two $d=3$ supermultiplets obtained
by reducing the $d=4$ scalar supersingleton can be regarded as
``supercurrents", i.e., as the squares of the $d=3$ supersingletons. The same
is true for any $N=2n$. For instance, in the case $N=4$ there are two types of
such $d=3$ supersingletons, differing by the SO(4) assignments of the scalars
and the spinors. The supersingleton corresponding to the supermultiplet
(\ref{1.18}) is described by the following ultrashort analytic superfield:
\begin{equation}\label{1.19}
  \F^{[+]} = \o_a u^{[+]}_a +
  \q^{[+]\{-\}\a} \chi_{\a \dot a} w^{\{+\}}_{\dot a} -
  \q^{[+]\{+\}\a} \chi_{\a \dot a} w^{\{-\}}_{\dot a} +  \q^{[+]\{+\}}\s^m\q^{[+]\{-\}}i\pa_m \o_a
  u^{[-]}_a\,,
\end{equation}
where $\square\o_a = \pa^{\a\b}\chi_\b = 0\,.$  Then the ``supercurrent"
(\ref{1.18}) is simply the square of the supersingleton (\ref{1.19}), $J^{[++]}
= (\F^{[+]})^2$. The alternative supersingleton $\F^{\{+\}}$ has the mirror
SO(4) assignments and gives rise to the mirror supercurrent $J^{\{++\}} =
(\F^{\{+\}})^2$. This realisation of the supermultiplet (\ref{1.18}) explains
why the vector $j_m$ is conserved: It is simply the current $j_m = i\o^a\pa_m
\o_a + \chi^{\dot a}\s_m \chi_{\dot a}$.

The generalization to any $N=2n$ is straightforward. The supermultiplet $J^{[0
\cdots 0 20_n]}$  can be viewed as a ``supercurrent", i.e., as the square of
the corresponding $d=3$ supersingleton
\begin{equation}\label{1.26}
  \F^{+_1\cdots+_{n-2}[+]} = \o_a u^{+_1\cdots+_{n-2}[+]}_a +
  \left(\q^{[+]\{-\}\a}\chi_{\a\dot a}w^{+_1\cdots+_{n-2}\{+\}}_{\dot a} +
  \mbox{perm.}\right)
  + \mbox{der. terms}
\end{equation}
$$\Rightarrow\quad  J^{[0 \cdots 0 20_n]} \equiv J^{++_1\cdots++_{n-2}[++]} =
\left(\F^{+_1\cdots+_{n-2}[+]} \right)^2$$ (and similarly for the mirror
multiplet).

\section{Reducing the F(4) supersingleton to SU(2,2/1) multiplets}

As explained in Section 2, the unique $d=5$ superconformal algebra F(4) can be
reduced to the $d=4$ $N=1$ superalgebra SU(2,2/1). In this case we find that
the $d=5$ scalar supersingleton (``hypermultiplet") is reduced to a pair of
$d=4$ scalar supermultiplets with ``anomalous" dimension 3/2. Here are the
details.

The $d=5$ pseudo-Majorana spinors $\q^i_\a = \O_{\a\b}\e^{ij}(\q^j_\b)^*$ carry
two kinds of indices: $\a=1,\ldots,4$ is a spinor index of USp(2,2) $\sim$
SO(4,1) and $i=1,2$ is a doublet index of SU(2). The scalar supersingleton is
very similar to the $d=4$ $N=2$ hypermultiplet described in Section \ref{HM}.
Introducing the same SU(2)/U(1) harmonic variables and projecting $\q^{1,2}_\a
= \q^i_\a u^{1,2}_i$, we can write down the ultrashort analytic superfield
\begin{equation}\label{1.08}
  W^1(x_A,\q^1,u) = \f^i(x_A) u^1_i + \q^1_\a \j^\a(x_A) + i \q^1\g^\m \q^1 \pa_\m \f^i(x_A)
  u^2_i\,,
\end{equation}
where the component fields are massless,
\begin{equation}\label{1.08'}
  \square \f^i = \g^\m_{\a\b}\pa_\m \j^\a = 0\,.
\end{equation}
Correspondingly, they have the canonical dimensions
\begin{equation}\label{candim}
 \ell_\f=3/2\;, \qquad \ell_\j= 2\;.
\end{equation}
The space-time variables $x^\m_A$ in (\ref{1.08}) are obtained from the real
ones $x^\m$ by a nilpotent shift, $x^\m_A = x^\m +i\q^1\g^\m\q^2$. The $d=5$
gamma matrices have the properties $\g^\m_{\a\b} = -\g^\m_{\b\a}$,
$\O^{\a\b}\g^\m_{\a\b} = 0$.

The reduction to $d=4$ is obtained by splitting the four-component USp(2,2)
spinor $\q^{1}_\a$ into left- and right-handed spinors of SL(2,$\mathbb{C}$),
$\q^{1}_\a \ \rightarrow\ \q^+_\a,\ \bq^+_{\da}$, both carrying charge $+1$
under the $d=4$ R symmetry group U(1). \footnote{Note that in Section \ref{HM}
the odd variables were vectors of SO(2) and carried charges $\pm2$, while here
they are treated as spinors of SO(2) with charges $\pm1$.} In order to obtain
an irreducible $d=4$ supermultiplet, we keep only, e.g., the left-handed half.
Further, the $d=5$ fields in (\ref{1.08}) become $d=4$ fields $\f^i u^1_i \
\rightarrow A^+$, $\j^\a \ \rightarrow \ \l^\a$ of non-canonical dimensions 3/2
and 2, respectively. The derivative term in (\ref{1.08}), when restricted to
left-handed $\q$'s only, gives rise to the scalar $B^-\equiv \pa_4\f^i u^2_i$
of shadow dimension $4-3/2=5/2$. Finally, the analytic basis variables $x^\m_A$
in (\ref{1.08}) become the left-handed chiral basis variables $x^m_L = x^m +
i\q^+\s^m \bq^-$. The end result is the $d=4$ chiral superfield
\begin{equation}\label{lchsup}
  J^+(x_L) = A^+(x_L) + \q^{+\a} \j_\a(x_L) + \q^{+\a} \q^+_\a B^-(x_L)
\end{equation}
with ``anomalous" dimension 3/2.

Alternatively, starting with the complex conjugate supersingleton and retaining
the right-handed halves of the odd variables, we can obtain the mirror $d=4$
chiral multiplet
\begin{equation}\label{rchsup}
  J^-(x_R) = A^-(x_R) + \bq^{-}_{\da} \bar\rho{\da}(x_R) +
  \bq^{-}_{\da} \bq^{-\da} B^+(x_R)\;.
\end{equation}
As before, the two pairs of shadow $d=4$ scalars $A^\pm,\ B^\pm$ needed to form
the two $d=5$ singletons are distributed over the two mirror supermultiplets.

The case $d=5\ \rightarrow\ d=4$ differs from $d=4\ \rightarrow\ d=3$ in that
the supermultiplets $J^\pm$ are not ``supercurrents", i.e., they are not
obtained as bilinears in the $d=4$ supersingletons (otherwise they would have
dimension 2 instead of 3/2).

\section*{Acknowledgements}
The work of S.F. has been supported in part by the European Commission RTN
network HPRN-CT-2000-00131 (Laboratori Nazionali di Frascati, INFN) and by the
D.O.E. grant DE-FG03-91ER40662, Task C. S.F. is grateful to M. Porrati for
useful discussions. E.S. acknowledges discussions with G. Arutyunov and A.
Petkou.


\begin{thebibliography}{99}

\bibitem{Fronsdal} C.~Fronsdal,
``Elementary Particles In A Curved Space. 4. Massless Particles,'' Phys.\ Rev.\
D {\bf 12} (1975) 3819.

\bibitem{FGPG} S.~Ferrara, R.~Gatto, A.~Grillo and G.~Parisi, ``The shadow
operator formalism for conformal algebra", Lett.\ Nuovo\ Cimento {\bf 4} (1972)
115.

\bibitem{DeWolfe:2001pq}
O.~DeWolfe, D.~Z.~Freedman and H.~Ooguri, ``Holography and defect conformal
field theories,'' arXiv:hep-th/0111135.

\bibitem{Porrati:2001db}
M.~Porrati, ``Higgs phenomenon for 4-D gravity in anti de Sitter space,''
arXiv:hep-th/0112166.


\bibitem{Klebanov:1999tb}
I.~R.~Klebanov and E.~Witten, ``AdS/CFT correspondence and symmetry breaking,''
Nucl.\ Phys.\ B {\bf 556} (1999) 89 [arXiv:hep-th/9905104].

\bibitem{ff2}
M. Flato and C. Fronsdal, {\it Lett. Math. Phys.} 2 (1978) 421; {\it Phys.
Lett.} 97B (1980) 236; {\it J. Math. Phys.} 22 (1981) 1100; {\it Phys. Lett.}
B172 (1986) 412.


\bibitem{Binegar:1982fv}
B.~Binegar, C.~Fronsdal and W.~Heidenreich, ``Conformal QED,'' J.\ Math.\
Phys.\  {\bf 24} (1983) 2828.



\bibitem{Gubser:1998bc}
S.~S.~Gubser, I.~R.~Klebanov and A.~M.~Polyakov, ``Gauge theory correlators
from non-critical string theory,'' Phys.\ Lett.\ B {\bf 428} (1998) 105
[arXiv:hep-th/9802109].

\bibitem{Witten:1998qj}
E.~Witten, ``Anti-de Sitter space and holography,'' Adv.\ Theor.\ Math.\ Phys.\
{\bf 2} (1998) 253 [arXiv:hep-th/9802150].


\bibitem{bgg}
I. Bars and  M. G\"unaydin, {\it Commun. Math. Phys.} 91 (1983) 31; M.
G\"unaydin and S. J. Hyun, {\it J. Math. Phys.} 29 (1988) 2367; M. G\"unaydin
and C. Saclioglu, {\it Commun. Math. Phys.}, 87 (1982) 159; {\it Phys. Lett.}
B108 (1982) 180.

\bibitem{ggg}  M.
G\"unaydin, {\it Oscillator Like Unitary Representations Of Noncompact Groups
And Supergroups And Extended Supergravity Theories}, Extended version presented
at 11th Int. Colloq. on Group Theoretical Methods in Physics, Istanbul, Turkey,
Aug 23-28, 1982. Published in Istanbul Grp.Th.Meth. (1982) 192; H. Nicolai,
{\it Representations Of Supersymmetry In Anti-De Sitter Space}. Published in
Trieste School (1984) 368; M. Duff, {\it  Int. J. Mod. Phys.} A14 (1999) 815.


\bibitem{Ferrara:1977sc}
S.~Ferrara, ``Algebraic Properties Of Extended Supergravity In De Sitter
Space,'' Phys.\ Lett.\ B {\bf 69} (1977) 481. 999) 815.

\bibitem{Nahm:1977tg}
W.~Nahm, ``Supersymmetries And Their Representations,'' Nucl.\ Phys.\ B {\bf
135} (1978) 149.


\bibitem{Kac}
V.~G.~Kac, ``Lie Superalgebras,'' Adv.\ Math.\  {\bf 26} (1977) 8.



\bibitem{Galperin:1984av}
A.~Galperin, E.~Ivanov, S.~Kalitsyn, V.~Ogievetsky and E.~Sokatchev,
``Unconstrained N=2 Matter, Yang-Mills And Supergravity Theories In Harmonic
Superspace,'' Class.\ Quant.\ Grav.\  {\bf 1} (1984) 469; ``Unconstrained
Off-Shell N=3 Supersymmetric Yang-Mills Theory,'' Class.\ Quant.\ Grav.\  {\bf
2} (1985) 155.


\bibitem{Howe:md}
P.~S.~Howe and G.~G.~Hartwell, ``A Superspace Survey,'' Class.\ Quant.\ Grav.\
{\bf 12} (1995) 1823; ``(N, p, q) harmonic superspace,'' Int.\ J.\ Mod.\ Phys.\
A {\bf 10} (1995) 3901 [arXiv:hep-th/9412147]; P.~Heslop and P.~S.~Howe, ``On
harmonic superspaces and superconformal fields in four dimensions,'' Class.\
Quant.\ Grav.\  {\bf 17} (2000) 3743 [arXiv:hep-th/0005135].

\bibitem{Galperin:uw}
A.~S.~Galperin, E.~A.~Ivanov, V.~I.~Ogievetsky and E.~S.~Sokatchev, ``Harmonic
Superspace,'' {\it  Cambridge, UK: Univ. Pr. (2001) 306 p}.


\bibitem{Ferrara:1998jm}
S.~Ferrara and C.~Fronsdal, ``Gauge fields as composite boundary excitations,''
Phys.\ Lett.\ B {\bf 433} (1998) 19 [arXiv:hep-th/9802126].

\bibitem{Gunaydin:vz}
M.~Gunaydin and N.~Marcus, ``The Unitary Supermultiplet Of N=8 Conformal
Superalgebra Involving Fields Of Spin <= 2,'' Class.\ Quant.\ Grav.\  {\bf 2}
(1985) L19.

\bibitem{Andrianopoli:1999vr}
L.~Andrianopoli, S.~Ferrara, E.~Sokatchev and B.~Zupnik, ``Shortening of
primary operators in N-extended SCFT(4) and  harmonic-superspace analyticity,''
Adv.\ Theor.\ Math.\ Phys.\  {\bf 3} (1999) 1149 [arXiv:hep-th/9912007];
S.~Ferrara and E.~Sokatchev, ``Superconformal interpretation of BPS states in
AdS geometries,'' Int.\ J.\ Theor.\ Phys.\  {\bf 40} (2001) 935
[arXiv:hep-th/0005151]; ``Conformal primaries of OSp(8/4,R) and BPS states in
AdS(4),'' JHEP {\bf 0005} (2000) 038 [arXiv:hep-th/0003051].












\end{thebibliography}
\end{document}